\documentclass[epj,final]{svjour}
\usepackage{graphicx}

\def\fv{{\bf{f}}}
\def\rv{{\bf{r}}}
\def\LA{\left\langle}
\def\RA{\right\rangle}
\def\LP{\left(}
\def\RP{\right)}
\def\LS{\left[}
\def\RS{\right]}

\begin{document}

\title{Microscopic Method for Dislocation Tracking}

\author{ Marco Patriarca \inst{1} \fnmsep \thanks{Corresponding author, Email: Marco.Patriarca@hut.fi}
    \and Miguel Robles\inst{2} \and Kimmo Kaski\inst{1} }

\institute{ Laboratory of Computational Engineering, 
and Research Center for Computational Science and Engineering,\\
Helsinki University of Technology, P.\,O. Box 9203, FIN--02015 HUT, Finland
\and        
UNAM, Centro de Investigacion en Energia,\\
Priv. Xochicalco S/N, Col. Centro, Temixco, Mor., A.P. 34, C.P. 62580, Mexico
}

\abstract{
We show that a microscopic definition of crystal defect, based on 
the effective mean single-particle potential energy, makes it possible 
to detect and visualize various types of local and extended crystal defects and
develop an effective algorithm for tracking their time evolution.
}

\maketitle

\section{Introduction}
\label{intro}

Real crystalline solids are characterized by the presence of various types
of defects, both local and extended.
Vacancies, interstitials, and impurities, are examples of local defects,
while extended defects can either be linear and have a one-dimensional (1D) character,
as in the case of dislocations, or have a two-dimensional (2D) character,
as in the case of grain boundaries.
Dislocations, in particular, are specifically interesting and complex 
entities with quite peculiar dynamical characteristics. 
On one hand, they are known to have a well defined identity,
which can be characterized at the mesoscopic or macroscopic scale. 
On the other hand they originate from and evolve driven by 
microscopic interactions, under the influence of thermal fluctuations.
This raises the question -- especially from the atomic level perspective --
of a proper definition for a dislocation.
A definition, which is based on the physical mechanisms 
responsible for their formation and evolution,
can be expected to be suitable for studying their general properties, such as
stability, dynamics, mutual interactions, and nucleation conditions.
Usually, dislocations in crystalline materials are defined geometrically
as topological defects, through the Burgers vector, 
which represents an invariant measure of the deviation of the structure of a 
real specimen from that of a corresponding perfect crystal~\cite{Hirth+Lothe}.
By ``invariant'' one means that the Burgers vector is independent 
on the contour used to compute it and that it is sufficient 
to completely specify the dislocation in the framework of elasticity theory.
However, from the computational point of view,
for example in Molecular Dynamics (MD) or Monte Carlo (MC) studies of crystals,
this kind of definition may be expensive if applied directly.
Furthermore, it would require a partitioning of the crystal into a ``bad'' 
part, close to the dislocation core, and a ``good'' part, far from it
and similar, in its local structure, to an ideal crystal~\cite{Hirth+Lothe}.
Quantitatively, this partition can turn out to be ambiguous or difficult 
to realize as a program, e.\,g. if the type and position of the dislocation is not known or 
other crystal defects are present.
Physical properties which characterize the nature of crystal locally
are expected to be better candidates for numerical studies of dislocations.

This paper focuses on the search of a simple and non-ambiguous criterion
for detecting and automatically tracking the time evolution of dislocations 
and other types of crystal defects, which can be helpful and computationally 
convenient in numerical studies of crystal dynamics.
Its aim is to overcome the difficulties typical of this kind of studies,
which concern detection of crystal defects as well as their visualization,
especially in 3D systems, where they are usually hidden by 
the bulk atoms and their topology is in general more complex than in the 2D case.
An automatic procedure for detecting and tracking crystal defects, illustrated
in the following, was developed from that used in the environment of the program
``Fracture'' for studying 2D crystals ~\cite{merimaa:99a} and then applied 
to dislocation dynamics in 2D systems~\cite{kuronen:01a,robles:02a}.

We can summarize the motivations for the present study in terms of a few
features that the desired tracking procedure is supposed to have.
First of all, an effective procedure should enable one to describe all types 
of crystal defects at the same time, not only dislocations, 
since in general different (types of) crystal defects 
can be present interacting with and transforming into each other. 
Secondly, it should not require detailed information about 
the position of the crystal defect to be studied.
Furthermore, the procedure should be efficient from the computational point of view, 
in order to be usable in large-scale MD and MC simulations of many-body systems 
out of equilibrium, like those usually met in the study of dislocations.
Finally, it should have a simple link with elasticity theory.

In Section~\ref{overview} we briefly review some techniques used
for detecting dislocations.
In Section~\ref{model}, after illustrating the model system and 
the main features of the numerical tools used for performing the simulations,
we describe the tracking procedure.
We show that properly defined mean single-particle potential 
energy provides a considerable amount of information about the lattice structure, 
which can be used to build an algorithm for the automatic tracking of dislocations.
Some examples and applications are provided in Section~\ref{examples},
ranging from the detection of point defects and grain boundaries
to the automatic tracking of the movement of dislocations.
Conclusions are drawn in Section~\ref{conclusion}.

\section{ Techniques for dislocation detection }
\label{overview}

While, in many cases, numerical simulations allow direct observation and analysis 
of the structure of a dislocation through comparison with the unperturbed lattice,
in the general case detecting a dislocation in a 3D system is no easy task,
if no previous information is available.
This can happen because computationally expensive 
work may be required for numerically evaluating the Burgers vector along 
several paths, before the dislocation is located exactly.
However, this can be avoided by characterizing dislocations in terms of local properties,
such as the atomic displacements from the sites of an ideal crystal,
rather than their equivalent integral properties, i.\,e., the Burgers vector.

In fact, some methods for detecting crystal defects use local properties,
defined in terms of the distances of the neighbor atoms, 
i.\,e., they measure the local strain~\cite{duesbery:98,vitek:70}
or the Burgers vector itself, but along small local circuits~\cite{schiotz:95},
or use information about the symmetry of the crystal studied~\cite{kelchner:98}.

Alternatively, the number of first neighbors $N_{\rm c}$, the coordination number, 
which can be obtained by the Voronoi tessellation or the Delaunay triangulation 
method~\cite{honneycutt:87,Okabe}, can represent an effective indicator of local order.
This tracking method is of a quite general nature and can be used to measure local order also 
in very disordered systems, such as glasses and liquids~\cite{bishop:2001,jund:97}.

However, in many problems one is interested in detecting and tracking only a few 
defects in an otherwise ordered crystal.
In these cases a tracking method based on dynamical microscopic, rather than geometrical, 
quantities, may be computationally convenient.
In the following we concentrate on a definition of dislocation based on the mean potential
energy felt by a particle, due to all the other particles of the system, and 
show that it is helpful in building computational routines for both practical 
recognition and movement tracking in atomic scale computer simulations.   
This choice can be considered to be convenient in that the potential energy is 
already computed by the basic algorithms of MD or MC codes,
so that no additional numerical procedures and computer time are required.

Potential energy is known to be a helpful parameter for visualizing dislocations,
both in 2D~\cite{altschuler:97,robles:02a} and 3D~\cite{schaaf:2000,zhou:98} systems.
The point of interest here is that potential energy-based methods can be used
to perform an automatic detection and tracking of dislocations~\cite{robles:01a,robles:02a}
during a numerical simulation, as illustrated below.
A potential energy-based method, which was shown to be able to automatically detect, 
count, and track the motion of dislocations in the 2D case~\cite{kuronen:01a,robles:02a},
is here generalized to the 3D case and used to study some examples of crystal defects.

\section{  Modeling and tracking of crystal defects }
\label{model}

\subsection{Computational tools}

Numerical simulations were carried out by using our code ``Boundary'',
which is a program based on coupling an MD code and a graphical user interface.
The original version of the code, developed for studying dislocations in
2D crystals~\cite{merimaa:99a,kuronen:01a,robles:02a},
was generalized to the 3D case with the purpose of studying general types of crystals defects.
In practice the code ``Boundary'' can be considered as a 3D molecular visualizer, 
coupled to and working in real time with an independent MD code.
Besides visualizing the system interactively, the program has some features 
which are particularly useful for the present investigation.
For instance, the most of the dynamical parameters and visualization modes can be varied 
during the numerical simulation and atoms can be visualized according to a color map 
based on different quantities, such as potential or kinetic energy, 
and selected according to the values of the same quantity.
In this way it is possible to search and detect crystal defects, 
which are not visible in physical space,
by exploring the system along additional dimensions, such as energy scales.

\subsection{Model system and units}

The model system considered in the following is a fcc specimen, 
made up of $N$ classical particles interacting with each other through a two-body 
Lennard-Jones (LJ) potential.
The interaction energy at time $t$ between two generic particles $i$ and $j$, 
at positions $\rv_i(t)$ and $\rv_j(t)$, respectively, is given by 
$W(|\rv_i(t) - \rv_j(t)|)$, where the LJ pair-wise potential reads
\begin{equation}
W(r) = 4\epsilon \LS \LP\frac{\sigma}{r}\RP^{12} - \LP\frac{\sigma}{r}\RP^6 \RS\ .
\label{LJ}
\end{equation}
A cut-off version of this potential was used to speed up the computation.
It was obtained by a constant shift in the energy scale 
and the addition of a term linear in $r$, in order to make 
both the potential and the force continuous functions of distance
-- see e.\,g. Ref.~\cite{merimaa:99a} for details.
Generalizations of the tracking procedure to other types of interactions, such as 
many-body empirical potentials and ab-initio potentials of Car-Parrinello type,
are straightforward, since the procedure is independent of the details of the interaction.
The LJ potential sets the parameters $\epsilon$ and $\sigma$
as natural energy and length scales, respectively.
In the following we use the standard LJ rescaled units, i.\,e.,
lengths are in units of $\sigma$, 
energies in units of $\epsilon$, and rescaled temperatures $T$ are defined as 
$T = k_{\rm B} T' / \epsilon$, where $T'$ is the physical temperature.
\begin{figure}
\centering
\includegraphics[width=3.2in]{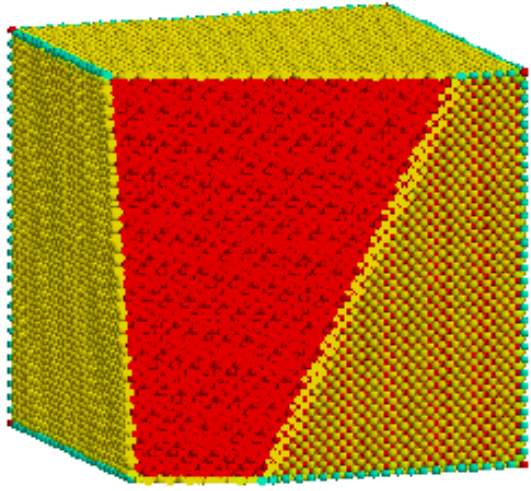}\\
\includegraphics[width=3.2in]{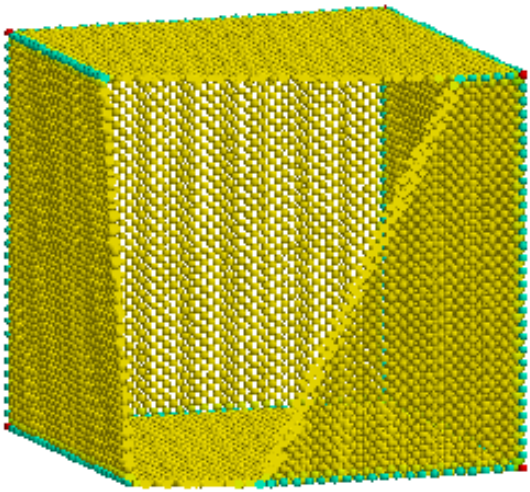}
\caption{
A model system at $T=0.05$ (top).
It is a cubic fcc specimen, with linear size of $30$ lattice constants, 
which approximately contains $10^5$ classical Lennard-Jones particles.
Atoms are colored according to potential energy.
Boundaries are visualized (bottom) by selecting atoms with high mean potential energies, 
corresponding to the last three peaks in Fig.~\ref{pe_spectrum}(b), see text for details.
The sample is cut to show its interior.
}
\label{specimen}
\end{figure}

\subsection{Single-particle potential energy}

We now turn to the definition of the mean single-particle potential energy $U_i$, felt by 
the $i$-th atom, which is the quantity used in the following
for monitoring the local structure of the crystal,
\begin{equation}
  U_i = \sum_{j=1 (j\ne i)}^{N} W_{ij}\ ,~~~~~~~~~~~i=1,\dots,N,
  \label{Ui}
\end{equation}
where $W_{ij} \equiv W(|\rv_i-\rv_j|)$ is the interaction energy between particles $i$ and $j$.
Even if the quantity $U_i$ is in principle a function of the $3N$ coordinates of all particles,
it nevertheless has some remarkable properties of a single particle quantity, if regarded as 
a function of $\rv_i$, for fixed $\{\rv_j\}$, $j \ne i$.
From the dynamical point of view, 
$U_i$ can be considered as the instantaneous effective potential 
felt by the $i$-th particle, since its gradient yields 
the total force $\fv_i$ acting on the $i$-th particle, due to the other particles,
\begin{equation} 
\fv_i = - \nabla_i U_i = - \sum_{j(\ne i)} \nabla_i W(|\rv_i-\rv_j|)\ .
\label{f}
\end{equation}
Furthermore, the halved sum of the $U_i$'s gives the total internal energy $U_{\rm tot}$,
\begin{equation} 
  U_{\rm tot} =  \frac{1}{2} \sum_{j \ne i} W_{ij} \equiv \frac{1}{2} \sum_{i} U_{i}\ .
\label{U}
\end{equation}
Thus, on a coarse-grained scale, $U(\rv_i)=U_i$ is a function of position $\rv_i$ 
proportional to the elastic potential energy density
and, therefore, directly related to the local deformations.

\subsection{Defect detection: The example of boundaries}

The distribution of atoms in potential energy has a very simple form in the case 
of an ideal (atomic) crystal.
At zero temperature all atoms have exactly the same energy $U_i \equiv U_0$ and
correspondingly the potential energy spectrum only consists of a single peak 
-- in fact a $\delta$-function -- located at $U=U_0$.
At temperatures larger than zero this peak will undergo a broadening,
due to atoms oscillating around their equilibrium positions.
One can then define a crystal defect, at least in the limit of zero temperature,
as a local or extended group of atoms producing some deviations 
in the ideal potential energy spectrum.

\begin{figure}
\centering
\includegraphics[width=3.2in]{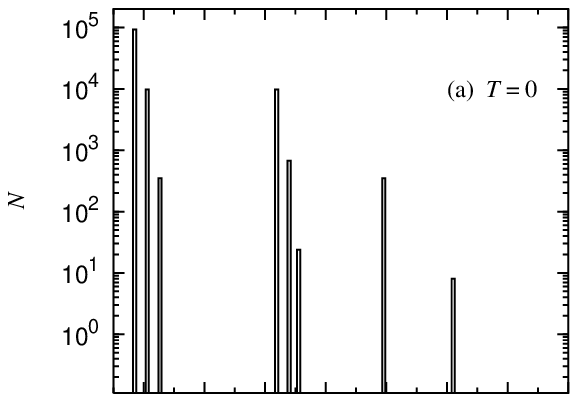}\\
\includegraphics[width=3.2in]{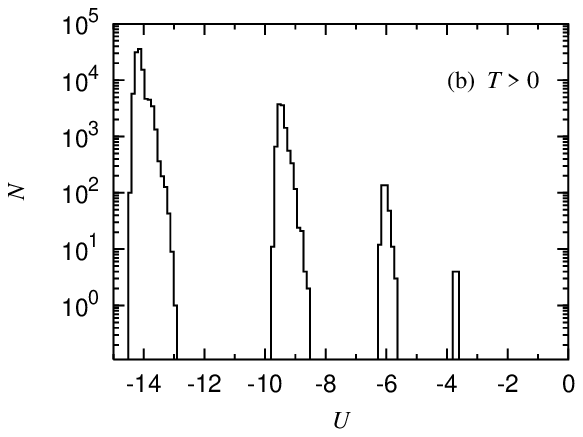}
\caption{
Potential energy spectra of the system at $T=0$ (a) 
and in thermal equilibrium at $T=0.05$ (b).
The potential energy spectrum can be resolved into four main peaks, 
which are due to bulk, surface, edge, and vertex atoms, from left to right, respectively.
}
\label{pe_spectrum}
\end{figure}

In order to study the potential energy spectrum,
we prepared the initial configuration by placing atoms on the sites 
of the perfect fcc lattice, corresponding to the interaction potential defined above,
with a rescaled cut-off $r_{\rm c}=2.5$.
The sample is shown in Fig.~\ref{specimen}.
It had a cubic shape, with sides aligned along the principal directions 
$[100]$, $[010]$, and $[001]$, and free boundary conditions were used.
The linear sizes were of $30$ lattice sites and the system contained
approximately $N = 1.1 \times 10^5$ atoms.
The system was either relaxed by over-damped MD to find the equilibrium configuration
at $T=0$ or thermalized, by using a Langevin thermostat, in order to reach a state of
thermal equilibrium at a temperature $T>0$.
The potential energy spectra of the system in equilibrium, at $T=0$ and $T=0.05$,
are compared in Fig.~\ref{pe_spectrum}.
One can notice that they have several maxima.
In this case the crystal defects which produce the deviations from 
the the ideal (single-peaked) spectrum are the boundaries of the system.
In fact, boundaries are a particular type of crystal defects, 
related to the finite size of the sample,
since they interrupt the infinite periodic structure of an ideal crystal.

The structure of the spectrum at zero temperature, shown in Fig.~\ref{pe_spectrum}(a),
can be briefly illustrated as follows.
The various maxima can be grouped into four main peaks.
From left to right, we will refer to them as the bulk, surface, edge, and vertex peaks, 
respectively, depending on the atoms they represent.
Among all these maxima, only the first maximum, of the three ones in which the first bulk peak 
can be resolved, corresponds to the ideal crystal. 
It is due to those atoms in the inner parts of the system, 
where the local crystal structure more closely resembles that of an ideal crystal,
which are further than the cut-off radius from the boundaries.
All other maxima are related to the presence of boundaries.
Namely, they are due to the different number of interacting neighbors that an atom can have, 
depending on its position in the sample.
For the given reduced cut-off radius $r_{\rm c}=2.5$,
some atoms, located in atom layers inside the bulk, close enough to the surface, 
interact with less neighbors than the other bulk atoms.
Correspondingly, they have a slightly higher mean potential energies and
produce the central maximum in the bulk peak in Fig.~\ref{pe_spectrum}(a).
Analogously, some other bulk atoms are placed inside the bulk in rows 
running very close to and along the edges.
They have even less neighbors and higher potential energies,
thus producing the third maximum of the bulk peak.
Analogous boundary effects take place for the surface atoms,
which lead to the splitting of the surface peak into three maxima.
The correspondence between the maxima and the boundary atoms
can be verified by scanning the potential energy spectrum through 
a window, which is narrow enough to selectively visualize atoms of a single maximum at time.
Furthermore, it is worth studying the dependence of the spectrum on the number of neighbors 
by varying the value of the cut-off.
When the cut-off is decreased one observes the maxima in the same peak to get closer,
until they merge in a single maximum, when atoms only interact with their first neighbors.
On the other hand, increasing the value of the cut-off leads to the appearance 
of many additional maxima.

At $T>0$ atoms oscillate around their equilibrium positions,
causing a broadening of all the maxima.
In the spectrum of the system in thermal equilibrium at $T=0.05$, 
shown in Fig.~\ref{pe_spectrum}(b), peaks cannot be resolved anymore into separate maxima.
The main peaks, however, can still be distinguished from each other quite clearly.

It is to be noticed that, while the bulk-peak is expected to be similar
for any samples of the same material, 
the structure of the other potential energy peaks depend on several features of the system.
For instance, the spectra in Fig.~\ref{pe_spectrum} corresponds to surfaces
perpendicular to the principal crystal directions $[001]$, $[010]$, and $[001]$, 
but different ways of cutting the crystal lead to different structures 
of the potential energy spectrum, which reflect the different surface energies.
Furthermore, boundary conditions can change the number of peaks appearing in the spectrum.
For example, if periodic boundary conditions are applied to the $x$- and $y$-dimensions,
the only boundaries will be the upper and lower surfaces perpendicular to the $z$-axis.
There will be no edges or vertexes and, correspondingly, the last two peaks on the right 
in the histograms in Fig.~\ref{pe_spectrum} will disappear.
Also the dependence on the dimensionality of the system should mentioned,
e.\,g. a 2D crystal has a 2D bulk, while boundaries are edges and vertexes.

An important point for the following considerations is the correspondence between the topology 
of crystal defects and the structure of the potential energy spectrum.
This was illustrated here for the particular cases of boundaries, 
considered as an example of defects which can have different dimensionalities, 
ranging from $D=0$ for point defects such as vertexes, 
to $D=1$ and $D=2$, for extended defects such as edges and surfaces, respectively.
This correspondence also holds for the other types of crystal defects and can be used 
for tracking and visualization purposes.
As a simple example of selective defect visualization, 
Fig.~\ref{specimen}(b) shows the system boundaries -- bulk atoms were excluded -- 
by selection of the atoms corresponding to the three peaks at higher potential energy,
in the spectrum in Fig.~\ref{pe_spectrum}(b).

\subsection{Defect tracking}
\label{tracking}

The same criterion, which allows one to detect crystal defects and selectively 
visualize the corresponding atoms, can be used to build a numerical procedure 
which automatically tracks defect positions and records the time evolution 
of their shape and structure.
The tracking procedure works in the following way.
First it carries out a selection of those atoms, which have potential energy 
within a suitable potential energy window,
in order to exclude all bulk atoms in zones where the crystal structure is almost perfect.
In the case of dislocations, one can use a lower threshold value $U_{\rm a}$ 
slightly higher than the bulk energy $U_0$, which can be computed as the average energy 
of the first peak at low potential energies~\cite{robles:02a}.
A value $U_{\rm a} \approx U_0 + nT$, where $n$ is a small integer and 
$nT$ is an estimate of energy fluctuations around the bulk value $U_0$, 
works well for small enough temperatures.
An additional upper threshold $U_{\rm b}$ can be used in order to exclude atoms 
in the boundaries.
This first step is usually sufficient to effectively visualize and track a moving dislocation.

Secondly, a cluster algorithm can be applied to the selected atoms, in order to 
identify crystal defects as clusters of high potential energy particles in physical space.
This can be useful e.\,g. to track the position of the dislocation core
in studies of dislocation dynamics.

At higher temperatures, thermal fluctuations generate short-lived defects,
which are spontaneously re-absorbed by the crystal after a very small number of time steps.
In order to prevent the tracking procedure from automatically detecting also 
short-lived defects, we replaced the mean potential energy $U_i(t)$ at time $t$
with its time average $\LA U_i(t) \RA = \tau^{-1} \int \, dt\, U_i(t)$,
where the integration is carried out numerically from time $t-\tau$ to time $t$.
The value of $\tau$, however, has a natural lower limit in the average life-time 
of a short-lived defect, in order that fluctuations to be effectively average out, while,
on the other hand, it should not be larger than the time scale on which the dislocation
moves of one lattice constant or changes appreciably, 
otherwise its evolution may be lost during the time average.
Also, the cluster algorithm can be helpful in distinguishing short-lived defects, 
which usually are due to some isolated atoms, 
from dislocations, whose core is an atom cluster with a 1D character.

\begin{figure}
\centering
\includegraphics[width=3.2in]{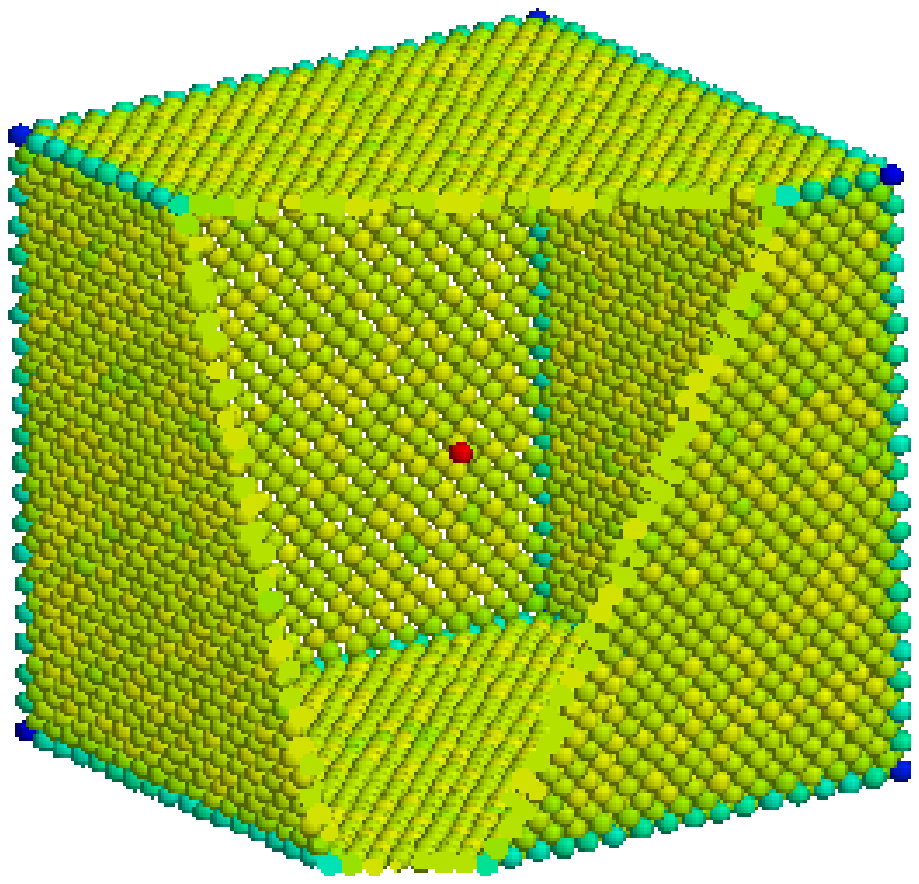}\\
\includegraphics[width=3.2in]{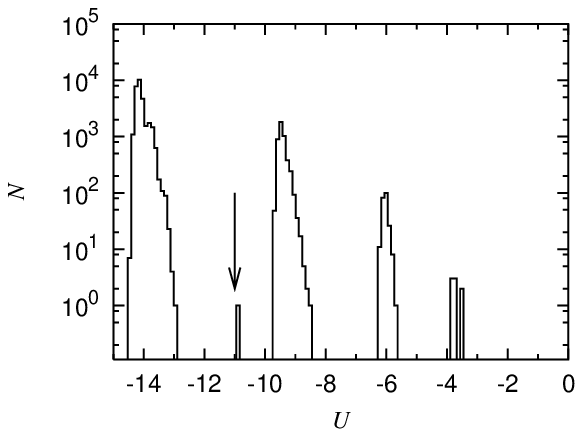}
\caption{
A substitutional impurity is placed in the middle of a cubic sample
with linear size of $20$ lattice constants. 
It is visualized (top) by selection,
in the potential energy spectrum (bottom),
of the corresponding peak (pointed by an arrow).
The sample was cut to show its interior.
}
\label{impurity}
\end{figure}
\begin{figure}
\centering
\includegraphics[width=3.2in]{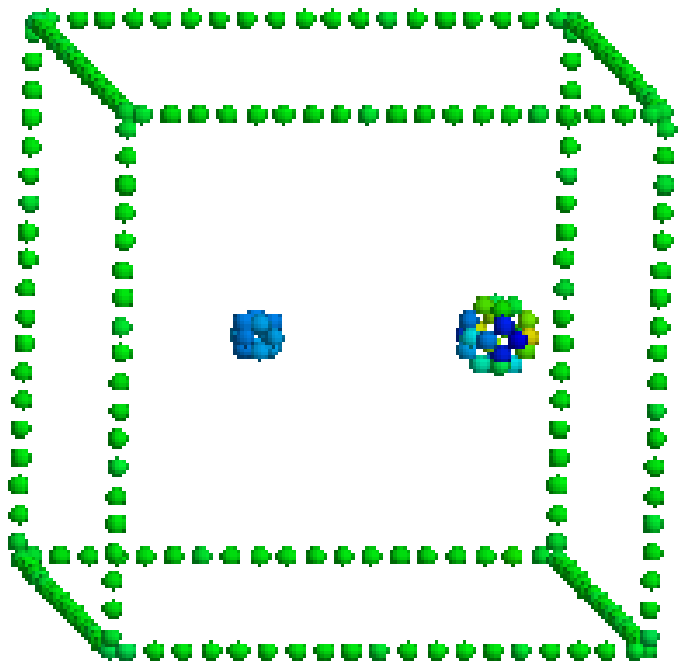}\\
\includegraphics[width=3.2in]{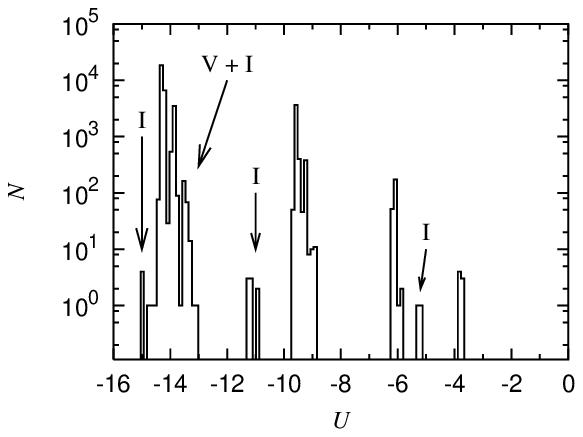}
\caption{
A sample system with a vacancy and an interstitial.
It is visualized (top) by selecting the atoms within the window $-13.8<U_i<-13.0$ of 
the potential energy spectrum (bottom).
Also some bulk atoms are shown, running along the system edges,
which are within the same potential energy window.
Label ``I'' indicates peaks originated from the interstitial.
The peak labeled ``V+I'' is due to an overlap of the vacancy peak and
one of the interstitial peaks.
}
\label{vacint}
\end{figure}
\section{Applications}
\label{examples}

\subsection{Point defects: impurities, vacancies, and interstitials}

Substitutional impurities have different potential energies, with respect to the potential
energy of the rest of the bulk atoms, due to the differences in the basic type of interaction.
On the other hand, an interstitial or a vacancy are characterized by 
a different number of first neighbors.
In all these cases, the presence of a point defect produces new peaks 
in the potential energy spectrum. 

As a first example of crystal defect detection, 
we consider a model of impurity, represented as a particle which interacts
with the other atoms of the system through a different LJ potential, 
characterized by parameters $\epsilon$ and $\sigma$ about $10\,\%$ larger and smaller, 
respectively, than those of the interatomic potential of the rest of the atoms.
The corresponding potential energy spectrum contains an additional peak, 
placed between the first (bulk) and the second (surface) peaks,
shown in Fig.~\ref{impurity}(b) indicated by an arrow.
By selecting a potential energy window $U_i>U_{\rm a}=-12$, one includes the impurity peak,
as well as the rest of the atoms at higher potential energies, while excluding bulk atoms
at the same time.
In this way one obtains Fig.~\ref{impurity}(a), 
in which the impurity is clearly visualized inside the specimen.

As another example, Fig.~\ref{vacint} shows the sample system, after one atom was moved 
from its original lattice site (where a vacancy was created) to a new position
(where an interstitial appeared).
After thermalization, there is a vacancy and a split-interstitial in the sample
and several new peaks appear in the potential energy spectrum, shown in Fig.~\ref{vacint}(b). 
One of the peaks -- that labeled ``V+I'' -- is due both to some atoms in a shell 
around the vacancy and some others around the interstitial.
The other additional peaks -- labeled ``I'' -- are due to atoms displaced
around the interstitial and the interstitial itself.
Figure~\ref{vacint}(a) was obtained by selecting atoms in the window with
$U_{\rm a}=-13.8$ and $U_{\rm b}=-13.0$,
which includes atoms both from the surroundings of the vacancy and of the interstitial.

\subsection{Extended defects: Edge and screw dislocations}
\label{dislocation}

As an example of detection and tracking of an extended crystal defect, 
we consider screw and edge dislocations in fcc crystals.
This type of dislocation is unstable in general, 
depending on the direction of its dislocation core.
Here we consider the case of a screw or edge dislocations with Burgers vector
along the $[110]$ direction, which is energetically unstable with respect 
to its splitting into two Shockley partials, 
according to the following reaction~\cite{Hirth+Lothe,Nabarro},
\begin{equation}
\frac{1}{2}[1 1 0] \rightarrow 
\frac{1}{6}[2 1 1] + \frac{1}{6}[1 2 \overline{1}]\ .
\label{splitting}
\end{equation}
The two partials are expected to initially move away from each other under the action 
of the mutual repulsion, until they either stop at their equilibrium 
distance $d$ or, if $d$ is larger than the sample size, reach the boundaries.
While the two partials are moving, a stacking fault ribbon remains between them.
We simulated both screw and edge dislocations by MD and found a very similar behavior.
In the rest of this subsection we concentrate on the case of the screw dislocation.

We prepared the initial configuration of the system by displacing atoms according 
to the analytical formula provided by elasticity theory~\cite{Hirth+Lothe,Nabarro},
\begin{equation}
u_{z} = \frac{b}{2\pi} \tan^{-1}\LP\frac{y}{x}\RP\ .
\label{uz}
\end{equation}
Here $z$ is the displacement along the core direction $[110]$, 
$x$ and $y$ are two (arbitrary) directions perpendicular to $z$ and to each other, 
while the modulus of the Burgers vector $b$ is given by $a/\sqrt{2}$, 
with $a$ equal to the lattice constant of the fcc crystal.
\begin{figure*}
\includegraphics[width=2.2in]{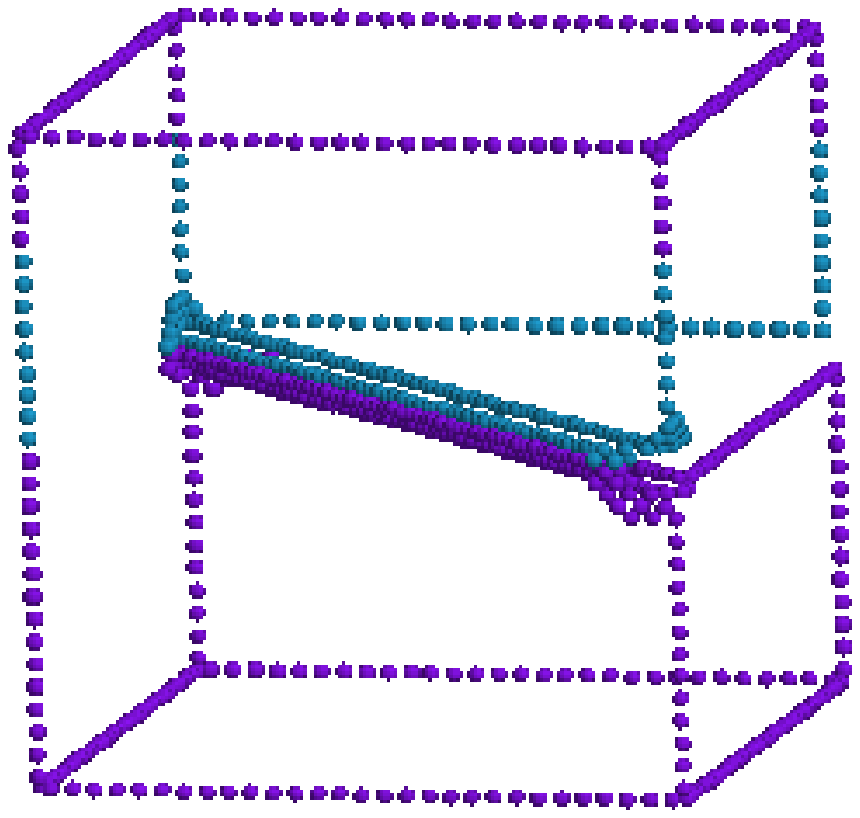}
\includegraphics[width=2.2in]{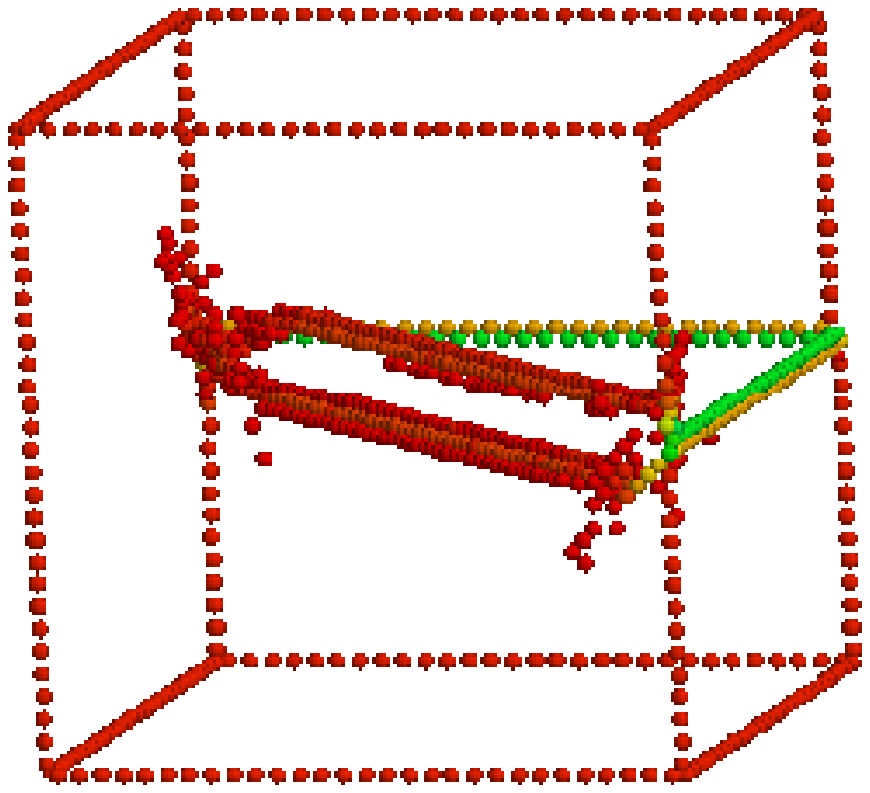}
\includegraphics[width=2.2in]{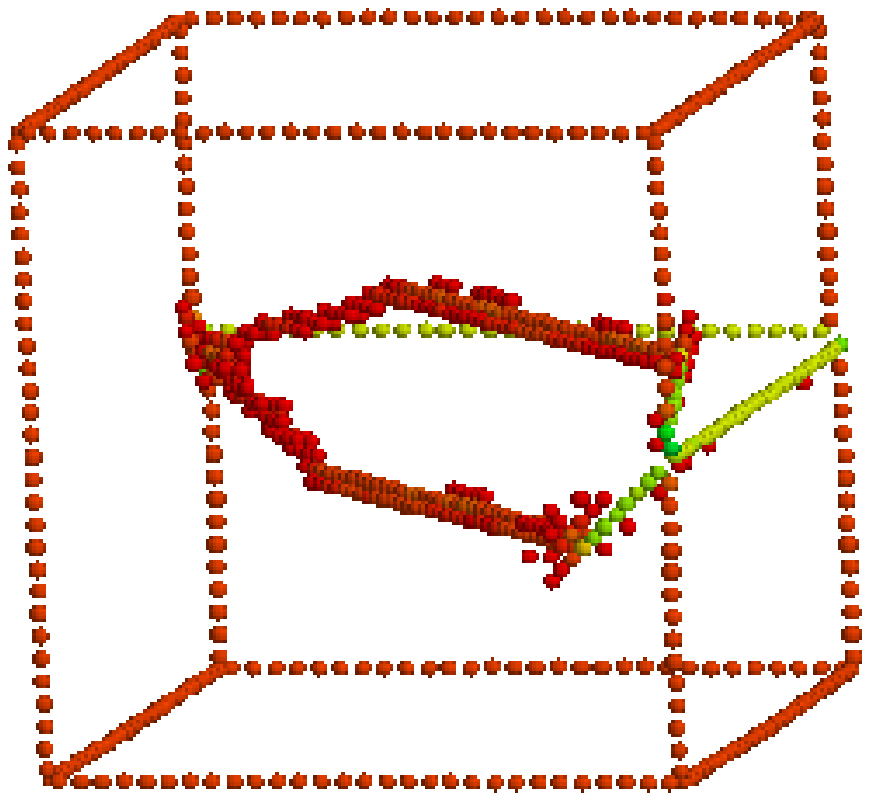}\\
\includegraphics[width=2.2in]{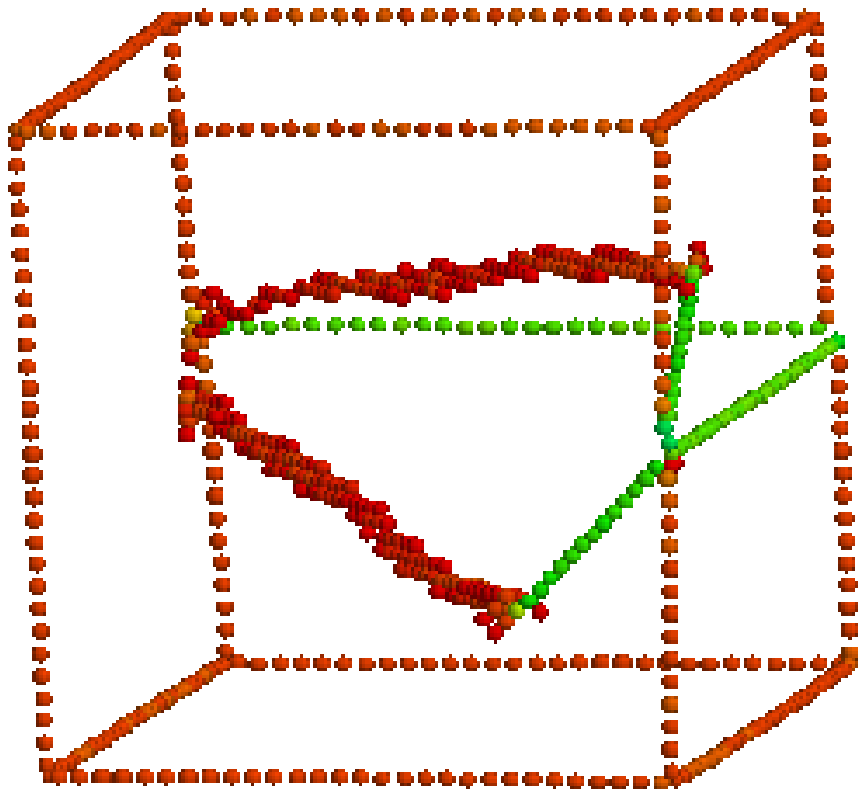}
\includegraphics[width=2.2in]{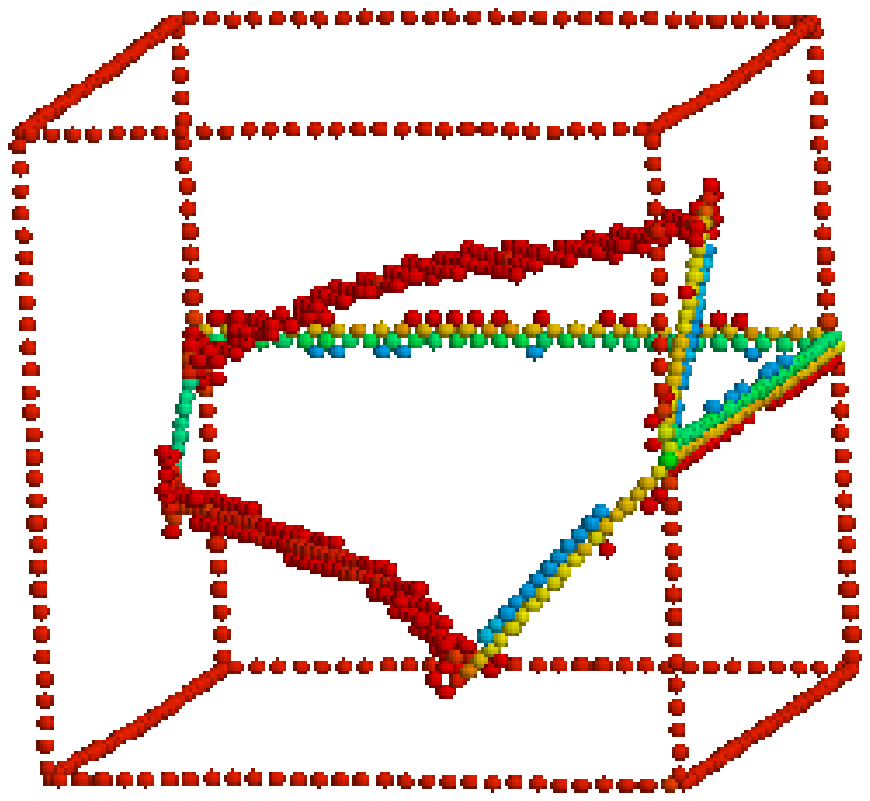}
\includegraphics[width=2.2in]{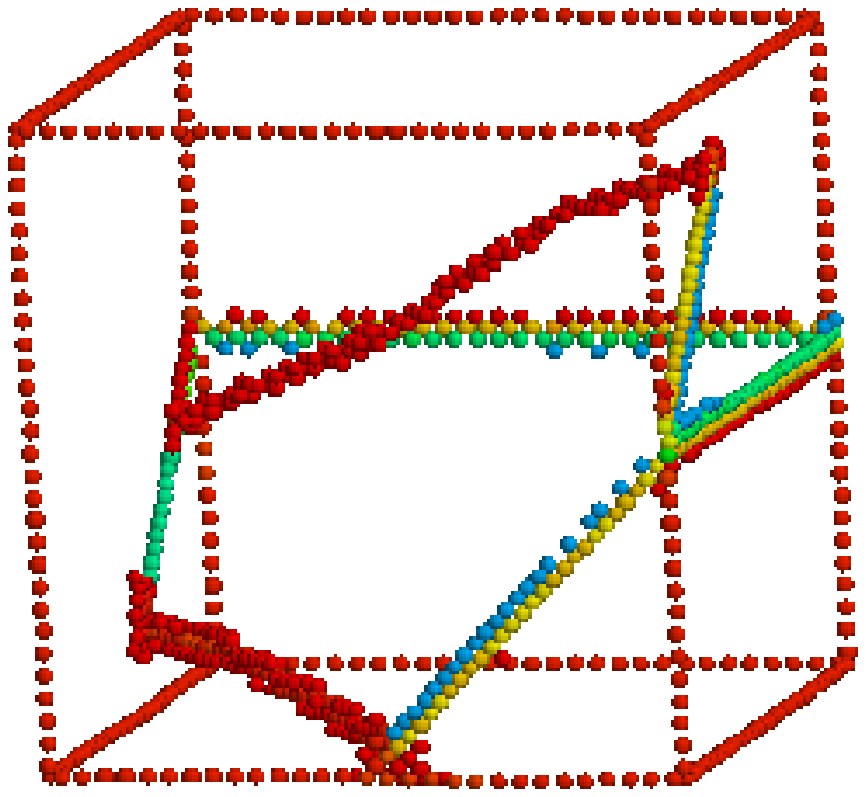}\\
\includegraphics[width=2.2in]{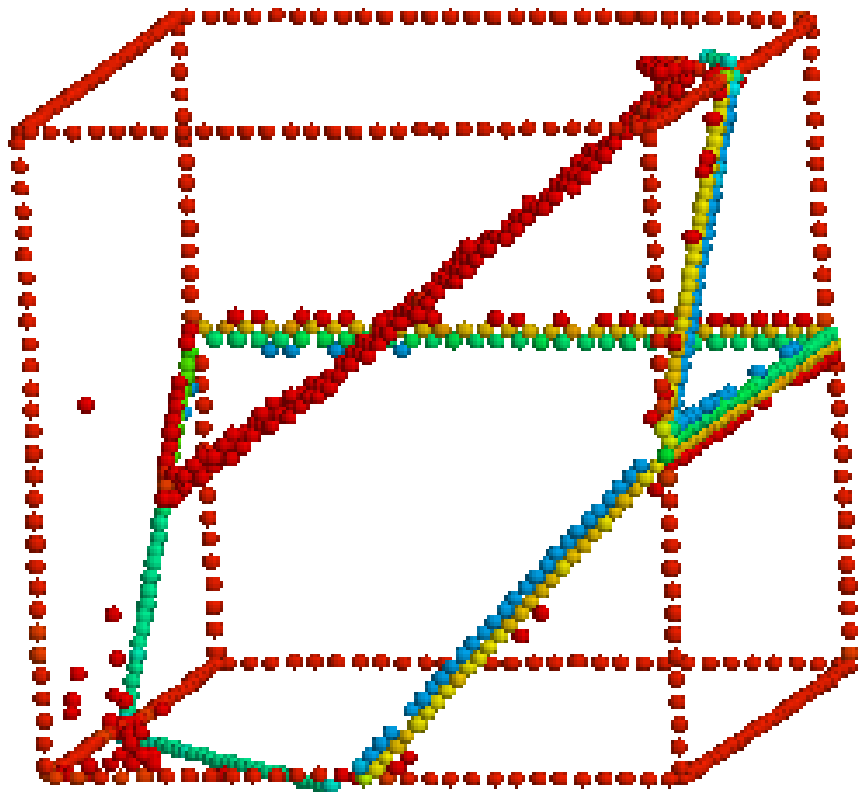}
\includegraphics[width=2.2in]{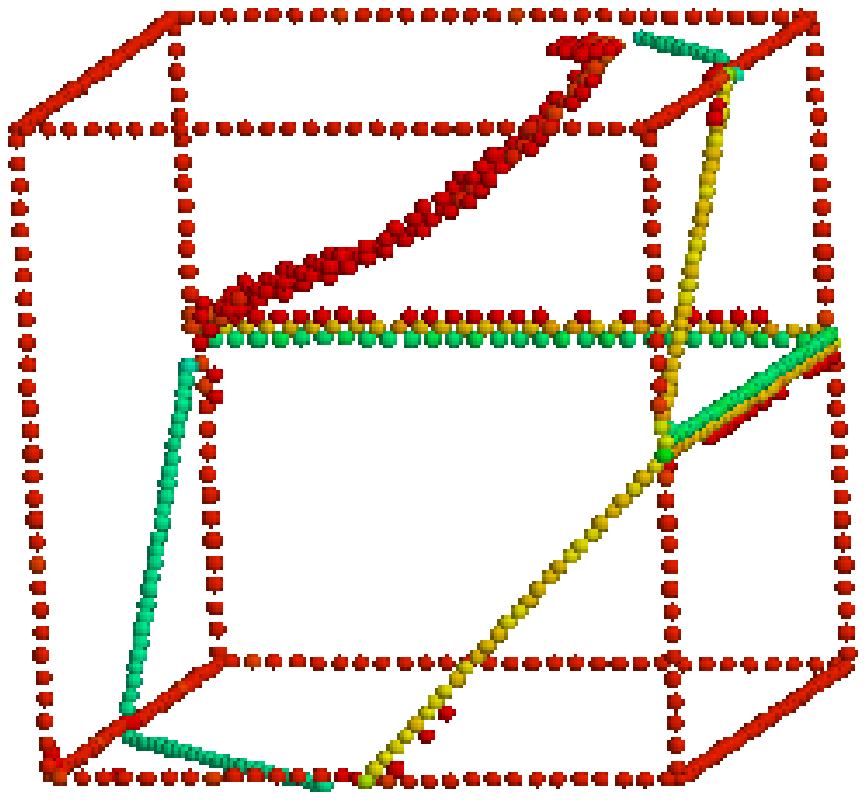}
\includegraphics[width=2.2in]{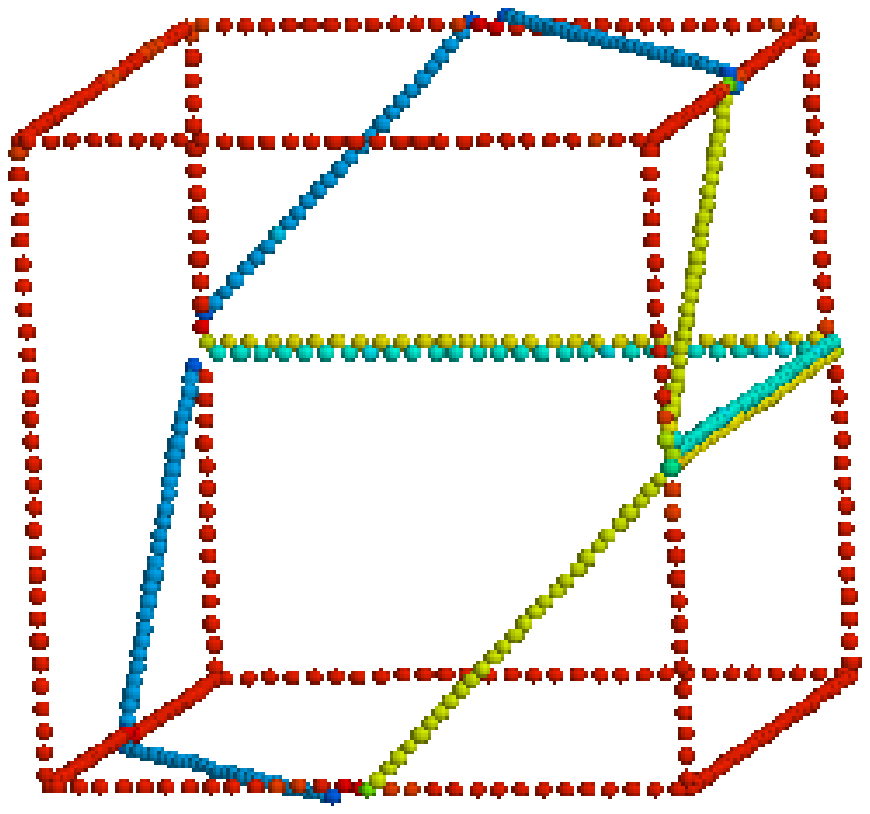}
\caption{
Time evolution of a screw dislocation splitting into two Shockley partials. 
The initial configuration (top-left) was prepared by displacing atoms according 
to linear elasticity theory, Eq.~(\ref{uz}).
The core of the initial dislocation is in the $[110]$ direction.
Only atoms with time averaged potential energies $-13.8<\LA U_i \RA<-12.0$ are shown
(apart from the first snapshot where slightly different values are used). 
In the equilibrium configuration (bottom-right), after the partials reached the boundaries
and disappeared, the intersection of the stacking fault, in the $(1 \overline{1} 1)$ plane,
with the surface of the cubic sample, produces a step with a hexagonal shape.
}
\label{screw}
\end{figure*}
In Fig.~\ref{screw} the initial configuration at $t=0$ is shown in the top-left snapshot.

\begin{figure}
\includegraphics[width=3.2in]{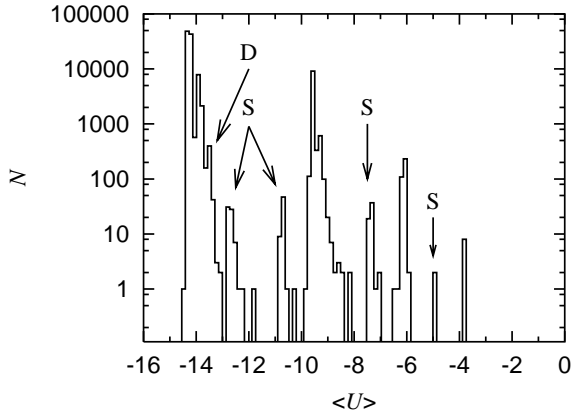}\\
\caption{
Potential energy spectrum of the sample with a screw dislocation.
Some additional peaks are present, compared with the spectrum of the unperturbed system
in Fig.~\ref{pe_spectrum}(b).
Only one of them -- labeled ``D''-- is due to the screw dislocation.
All the others --labeled ``S''-- are due to the various steps formed on the surface, 
either by the stacking fault or the initial deformation.
}
\label{screwS}
\end{figure}

The time evolution of the screw dislocation was visualized
by selecting atoms with time averaged potential energy within a suitable 
potential energy window, chosen as $U_{\rm a} = -13.8 < \LA U_i \RA < U_{\rm b} = -12.0$,
in the average potential energy spectrum shown in Fig.~\ref{screwS}. 
The potential energy $\LA U_i \RA$ was obtained by an average over an interval 
$\tau=0.1$ of ten time steps.
This selection includes atoms around the cores of the partial dislocations and, 
at the same time, eliminates both the bulk and the most of the boundary atoms.
There are also some other atoms, which are visible,
since their energies happen to be in the same potential energy window.
They are some atoms in the bulk, close to the edges,
and the atoms on two lateral sides, located on the steps which were formed 
at the initial time, when the initial screw dislocation was produced by displacing atoms.
The large strain introduced near the core of the dislocation, 
where Eq.~(\ref{uz}) is valid only approximately, 
causes a sudden increase of kinetic energy at the beginning of the simulation.
For this reason the system was thermalized during the very first time steps at $T=0.02$
and then evolved by constant energy MD.

The splitting of the screw dislocation into two partials was resolved almost immediately
during the very first steps of the time evolution.
In fact, the formation of a dislocation loop was observed.
Two edges of the dislocation loop remained parallel to the original core direction $[110]$,
while reducing in size.
The other two edges grew along the $[211]$ and $[12\overline{1}]$ directions, 
in agreement with Eq.~(\ref{splitting}).

\begin{figure}
\includegraphics[width=3.2in]{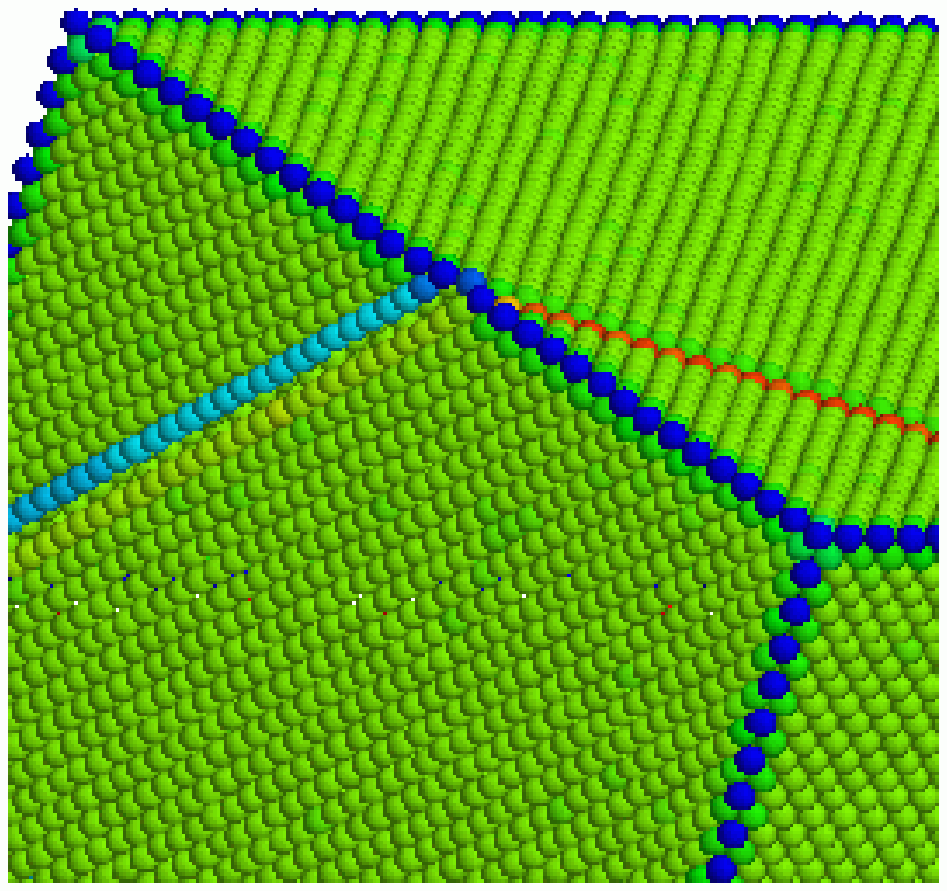}\\
\includegraphics[width=3.2in]{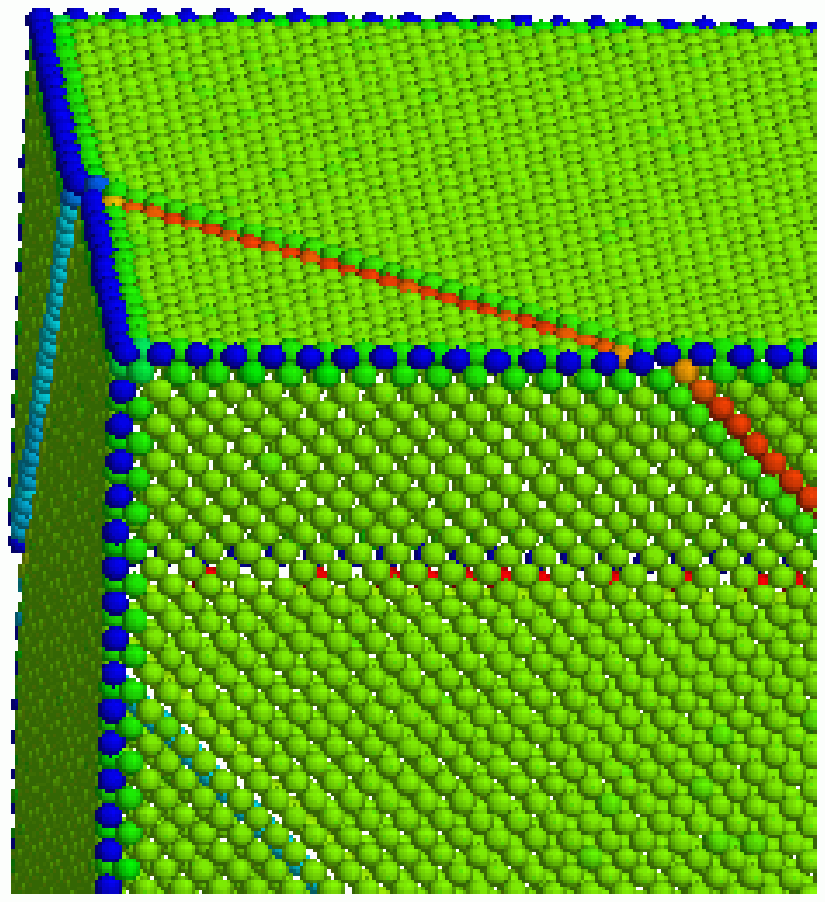}
\caption{
Enlarged details of the steps created by the intersection of the stacking fault
with the sample surface after the partials reached the surface and disappeared.
Atoms are colored according to their potential energies.
}
\label{step}
\end{figure}

During the time evolution of the dislocation loop, 
a stacking fault was formed inside the loop area.
In the particular crystal considered here, the stacking fault cannot be visualized directly 
through potential energy alone, since the corresponding atoms have no additional 
energy in relation to the perfect crystal structure.
However, its intersection with the sample surface produces some steps,
which are visible through the different potential energies of the atoms displaced.
One can notice that, initially, two small steps grew on the right side of the sample.
Then, when the whole partials became approximately parallel to the $[211]$ and 
$[12\overline{1}]$ directions, they began to translate 
and another step on the left side began to grow.
At equilibrium, after the partials reached the boundaries and disappeared,
the steps on the various surfaces, due to the stacking fault, formed a regular hexagon, 
due to the intersection of the $(1 \overline{1} 1)$ stacking fault plane 
with the cubic-shaped sample surface.
These steps are directly visible, 
as shown in the enlarged details in Figs.~\ref{step}(a) and (b).
\begin{figure}
\centering
\includegraphics[width=2.2in,angle=-90]{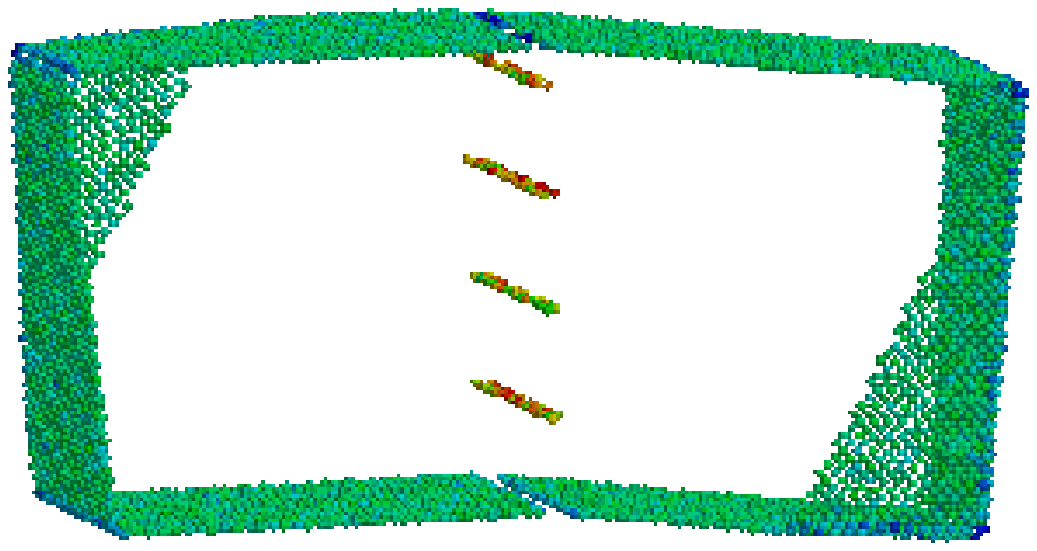}\\
\includegraphics[width=2.2in,angle=-90]{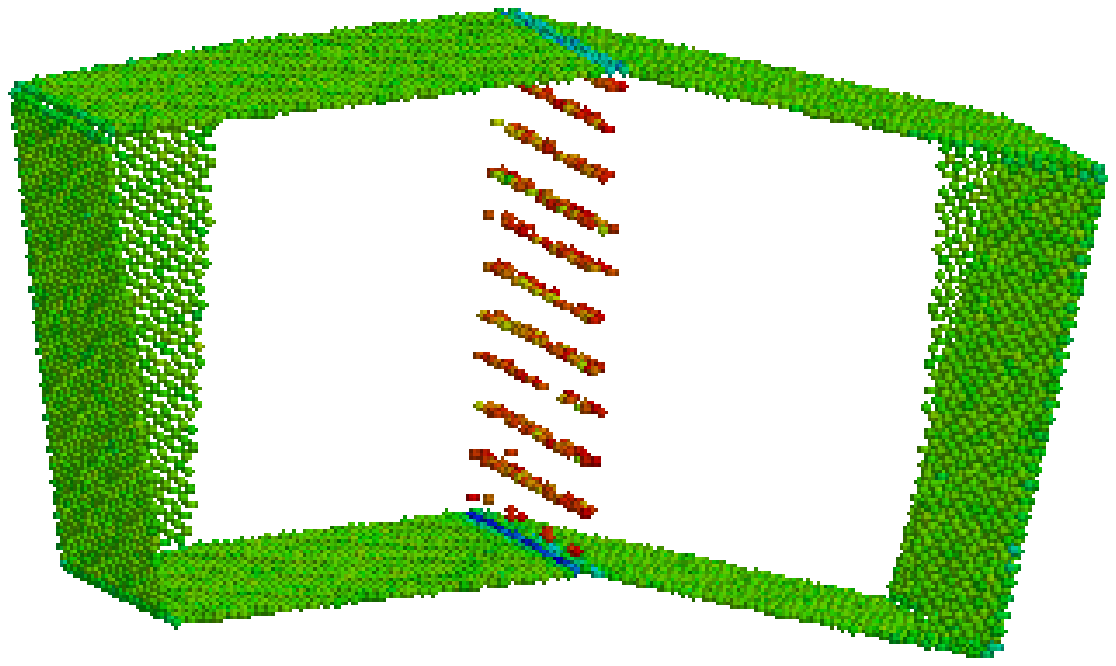}
\caption{
Symmetrical grain boundaries with angles $2\theta=8\,^{\circ}$ and $2\theta=20\,^{\circ}$, 
respectively, resolved in regular arrays of parallel dislocations.
Only atoms with potential energies $U_i>-10.7$ are shown.
The front and back boundary surfaces were removed.
}
\label{grain0}
\end{figure}

\subsection{Extended defects: Grain boundary}
\label{grain}

Grain boundaries are an example of planar extended defects, 
in that they represent the interface 
between two otherwise perfect crystal samples with arbitrary shapes and orientations.
Their general 2D nature is particularly evident in the case of nanocrystals, 
where grains can exist in a variety of sizes and types.
In the particular case of a low angle grain boundary, however,
they can be interpreted as arrays of 1D crystal defects
-- straight dislocations -- parallel to each other 
and lying in the grain boundary plane~\cite{Nabarro,read:50}.
This interpretation can be illustrated and verified by numerical experiment and
potential energy analysis, also for grain-boundaries with large angles, as follows.
We considered a system with an initial configuration obtained by joining 
two perfect cubic samples, each one of linear size of $30$ lattice constants.
The two twin samples were joined across the $(001)$ plane, in the $x$-direction.
Then one sample was tilted, with respect to the other one, of an angle $2\theta$, 
the tilt boundary lying along the $[010]$ direction.
The atoms, which overlapped as a consequence of the tilting, were removed.
After thermalization at $T=0.05$, the grain boundary 
was resolved into an array of dislocations parallel to the $[010]$ direction,
both in the case of $\theta = 4\,^{\circ}$ and in that 
of the larger angle $\theta = 10\,^{\circ}$, 
as shown in Fig.~\ref{grain0}.
Both Figs.~\ref{grain0}(a) and (b) were obtained by selecting high potential energy atoms 
with $U_i>U_{\rm a}=-10.7$.
The distance $\ell$ between dislocations is in agreement with the theoretical estimate, 
which gives a linear dislocation density $a/\ell = 2\sin(\theta/2)$~\cite{Nabarro,read:50}.
Namely, for $\theta = 4\,^{\circ}$ the spacing is either $7$ or $8$ lattice constants, 
the theoretical formula giving $\ell=7.17$, 
while for $\theta = 10\,^{\circ}$ we find a spacing of $2$ or $3$ 
lattice constants, to be compared with the theoretical value $\ell=2.88$.

\section{ Conclusions }
\label{conclusion}

We have shown that the mean potential energy felt by an atom,
due to all its neighbors, can be used as a tracer to detect and track
the time evolution of many types of crystal defects.
Usually the computation of potential energy is already carried out by MD or MC programs,
so that no additional computer time is required.
Despite the method cannot be applied at arbitrary temperatures and very disordered systems,
it is however possible to extend its validity to relatively high temperatures by using 
a time averaging and a cluster algorithm, together with the potential energy criterion.
The method illustrated here, based on the use of microscopic dynamical quantities,
can be considered to be complementary, rather than alternative, 
to other well verified methods based on geometrical approaches.
In this respect, through various examples, it was shown that such a microscopic approach
can be effectively exploited for describing crystal defects.

\begin{acknowledgement}

We thank Antti Kuronen, Peter Szelestey, and Michael Patra, for useful discussions.

This work was partially supported by the Academy of Finland,
Research Centre for Computational Science and Engineering,
project no.~44897
(Finnish Centre for Excellence Program 2000-2005).

\end{acknowledgement}

\end{document}